# A SHM method for detecting damage with incomplete observations based on VARX modelling and Granger causality

U. Ugalde[1], J. Anduaga[1], F. Martínez[1] and A. Iturrospe[2]

[1] *Department of Sensors, IK4-Ikerlan, Mondragon, Spain (uugalde@ikerlan.es)*
[2] *Department of Electronics and Computer Sciences, Mondragon Goi Eskola Politeknikoa, Mondragon, Spain*

*Abstract*—A SHM method is proposed that minimises the required number of sensors for detecting damage. The damage detection method consists of two steps. In an initial characterization step, substructuring approach is applied to the healthy structure in order to isolate the substructures of interest and later, each substructure is identified by a Vector Auto-Regressive with eXogenous inputs (VARX) model measuring all DOFs. Then, pairwise conditional Granger causality analysis is carried out with data measured from substructural DOFs to evaluate the information loss when measurements from all DOFs are not available. This analysis allows selecting those accelerometers that can be suppressed minimising the information loss. In the evaluation phase, vibration data from the reduced set of sensors is compared to the estimated data obtained from the healthy substructure's VARX model, and as a result a damage indicator is computed. The proposed detection method is validated by finite element simulations in a lattice structure model.

## 1 INTRODUCTION

Structural Health Monitoring (SHM) is a technological area which implements a damage detection and characterization strategies for engineering structures [1]. SHM is regarded as a very important engineering field in order to secure structural and operational safety; issuing early warnings on damage or deterioration, avoiding costly repairs or even catastrophic collapses [2].

Most of the existing vibration based SHM methods can be classified into two different approaches: global approaches and local approaches [3]. In the global approaches, the goal is to monitor the health of the entire structure. These global methods have been tested and implemented in different types of structures during the last 30-40 years [4]. However, for many large systems, global monitoring is not viable due to the lack of sensitivity of global features regarding local damages, inaccuracies of developed models, high cost of the computational operations etc [5]. On the other hand, local SHM methods are focused on evaluating the state of reduced parts within the entire structures by means of substructuring methods. This approach aims to overcome global method's limitations, dividing the whole structure into substructures and analysing each one individually.

Several papers have been published proposing substructuring methods for large scale structures. Koh et al [6] presented a "divide and conquer" strategy to monitor large structures based on the division of the whole structure into isolated substructures. For each substructure, structural parameters are identified using the extended Kalman filter (EKF). However, the EKF usually requires a prior analytical model, which is not always available [7]. Most recently, Xing et al [8] proposed another damage detection method using Auto-Regressive Moving Average with eXogenous input (ARMAX) models. The method is based on the extraction of natural frequencies from the ARMAX models and in their posterior analysis. Nonetheless it is preferably applicable to small and simple structures.

The authors of this article proposed in [9] a damage localization method based on the substructuring approach and Vector Auto-Regressive with eXogenous input (VARX) models. VARX models incorporate data from multiple DOFs and the method allows locating and quantifying damages within larger substructures. It is necessary to measure all substructural DOFs and consequently the amount of sensors that must be placed remain too high.

In general, if the number of sensors placed on a structure increases, the obtained information is more detailed. However, the high cost of data acquisition systems and accessibility limitations constrain in many cases the use of large number of sensors on a structure. Optimal sensor placement (OPS) methods aims to select a set of minimum number of sensors from all possibilities, such that the data collected can provide adequate information for the identification of the structural behaviour [10]. Different criteria can be used to evaluate the suitability of sensor locations, such as modal assurance criterion (MAC), measured energy per mode, information entropy (IE) or mutual information (MI).

Krammer [11] proposed an optimal sensor placement method based on modal properties. The measured mode shape vectors must be as linearly independent as possible in order to distinguish the measured or identified modes. MAC based sensor optimization methods aim to select points from all candidates to keep the original properties of the structure as well as it is possible.

Measured energy per mode criterion based OPS methods chooses those sensor positions with large amplitudes, which is critical in harsh and noisy circumstances. Heo et al. [12] focused on a kinetic energy to determine the optimal sensor placement on a long span bridge.

The two methods mentioned above are generally used when an analytical model (FEM) of the monitored structure is available. Nevertheless, if only experimental measurements are available, methods like the ones that are based on mutual information or information entropy can be used.

The mutual information (MI) of two random variables gives a measure of the mutual dependence between them and quantifies the "amount of information" obtained about one variable through the other variable. The optimal sensor locations can be obtained analysing which configuration gives the minimal mutual information values between sensors [13].

In SHM, it is desirable to design a certain sensor placement configuration that assures the measurement of representative data about the structural model parameters. The information entropy, as the measure of the uncertainty in the system parameters, gives the amount of useful information in the measured data. Information entropy theory is used to select the most informative sensors [14][15].

Granger causality (G-causality) is a statistical concept broadly applied in other scientific fields like neuroscience [16], climatology [17] or econometrics [18] with the purpose of analyse the causality among different time series. As mutual information and information entropy, does not require an analytical model and it is based on auto-regressive models.

In this paper, a SHM method is proposed for detecting damage with incomplete observations. Substructuring approach and VARX modelling are used to isolate and identify the healthy substructure. Pairwise conditional Granger causality analysis is carried out with data from all substructural DOFs to determinate which variables are the most proper ones to be observed. Damage is detected using the healthy VARX model and the data measured from the selected nodes.

The rest of the paper is organised as follows. In section 2, the proposed method is presented. In section 3, the method is evaluated by series of simulations and the obtained results are discussed. Finally the concluding remarks are exposed in section 4.

## 2 THE PROPOSED METHOD

A SHM method is proposed for detecting damage with incomplete observations. The substructuring approach is applied to isolate multi-DOF substructures from the healthy structure and each substructure is modelled by a corresponding VARX model as in [9]. In the healthy condition all DOFs of the substructure are measured and pairwise conditional Granger causality analysis is carried out with the data in order to determine which variables can be suppressed assuring the minimum worsening as it is possible. During the evaluation phase, a reduced set of data is measured again and by means of the healthy substructure's VARX model the condition of the substructure is evaluated.

*2.1. Substructure pairwise conditional Granger causality analysis*

Granger causality analysis measures the causal influence between stochastic processes and determines also the functional connectivity between them [19].

Traditionally, Granger causality analysis has been carried out between two stochastic processes. Assuming two jointly distributed stochastic processes ($X=X_1,X_2...X_n$ and $Y=Y_1,Y_2...Y_n$) it is said that $Y$ does not cause $X$ (according to Granger) if $X$ is only conditioned by its own past and therefore, $X$ is independent of the past of $Y$. Otherwise, if the past of $Y$ contains information about the future of $X$, it is said that $Y$ causes $X$ according to Granger.

More recently Barnett and Seth [20] developed a Multivariate Granger Causality (MVGC) toolbox, where time and frequency domain based multivariate G-causality approaches are available.

Within time domain multivariate G-causality approaches, pairwise conditional Granger causality is one of the most popular ones. In pairwise conditional G-causality, it is supposed that the universe $U$ is splits into three jointly distributed processes ($X$, $Y$, $Z$):

$$U = \begin{pmatrix} X \\ Y \\ Z \end{pmatrix} \quad (1)$$

The analysis wishes to eliminate any joint effect of $Z$ on the inference of Granger causality from $Y$ to $X$. Full and reduced VAR models are estimated from the multivariate processes and as is shown in equation 2 and 3, the difference between them is that variable $Y$ is not included in the reduced model:

$$X_n = \sum_{k=1}^{p} A_{xx,k} X_{n-k} + \sum_{k=1}^{p} A_{xy,k} Y_{n-k} + \sum_{k=1}^{p} A_{xz,k} Z_{n-k} + \varepsilon_{x,n} \quad (2)$$

$$X_n = \sum_{k=1}^{p} A'_{xx,k} X_{n-k} + \sum_{k=1}^{p} A'_{xz,k} Z_{n-k} + \varepsilon'_{x,n} \quad (3)$$

$A$ and $A'$ matrices correspond to the coefficients of the full and reduced VAR models, whereas $\varepsilon$ and $\varepsilon'$ are the residuals of both models. Furthermore, $p$ is the model order.

The causality from $Y$ to $X$ conditioned on $Z$ ($F_{Y \rightarrow X|Z}$) is stated as:

$$F_{Y \rightarrow X|Z} = \ln \frac{|\Sigma'_{xx}|}{|\Sigma_{xx}|} \quad (4)$$

where $\Sigma_{xx}$ and $\Sigma'_{xx}$ are the covariance matrices of the residuals of the full and reduced VAR models.

Figure 1 describes the proposed method for the selection of a certain substructural variable from a set of variables for its posterior suppression. In this method a G-causality value from each substructural variable to the whole set of substructural variables is analysed by a pairwise conditional G-causality approach. MVGC toolbox [20] is adapted to carry out the analyses.

The first step is the measurement of the displacements from the substructure ($U$). Once $U$ is measured, it is split into $X$, $Y$ and $Z$ variables. $X$ contains the displacement data that must be estimated, $Y$ contains the variables that their causality will be analysed and $Z$ holds the rest of measurements. As it is shown in figure 1, different variables are assigned to $X$, $Y$ and $Z$ in each iteration and the corresponding individual causality value ($F_{Y \rightarrow X|Z}$) is calculated. When all combinations have been analysed, the individual causality values that are calculated setting the same variable in $Y$ are added together in order to obtain a general causality value ($F_{G\ Y \rightarrow X|Z}$) for each variable.

The suppression of the variables with the lowest general causality values ($F_{G\ Y\rightarrow X|Z}$) cause the minimum worsening in the VAR model, so these variables are the most appropriate ones to be suppressed.

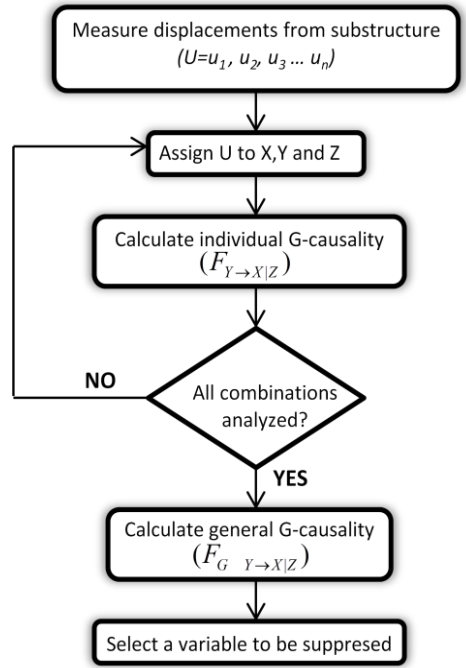

**Figure 1. Selection of variables for their suppression –flow chart**

## 2.2. Substructure damage detection with incomplete observations

A SHM method that aims to detect damages within substructures by means of incomplete measurements is described in this section. The method evaluates the condition of the structure using the VARX model of the healthy substructure and a set of variables that have been selected applying the approach described in the previous section.

We assume that the monitored substructure can be represented by the VARX model of equation 5. The VARX model contains $n$ endogenous variables ($y$) and $m$ exogenous variables ($x$), whereas $p$ and $q$ are the endogenous and exogenous model orders. $A_i$ and $B_i$ are endogenous and exogenous coefficient matrices.

We also assume that variable $y_1$ has been selected to be suppressed, so it is not measured never again during the evaluation phase. Thus, $y_2…y_n$ and $x_1, x_2… x_m$ are the measured displacements from the substructure, whereas $\hat{y}_1, \hat{y}_2…\hat{y}_n$ are the estimated displacement data.

$$\begin{bmatrix} \hat{y}_1(k) \\ \hat{y}_2(k) \\ ... \\ \hat{y}_n(k) \end{bmatrix} = A_1 \begin{bmatrix} \hat{y}_1(k-1) \\ y_2(k-1) \\ ... \\ y_n(k-1) \end{bmatrix} + A_2 \begin{bmatrix} \hat{y}_1(k-2) \\ y_2(k-2) \\ ... \\ y_n(k-2) \end{bmatrix} + ... + A_p \begin{bmatrix} \hat{y}_1(k-p) \\ y_2(k-p) \\ ... \\ y_n(k-p) \end{bmatrix} + \\ + B_1 \begin{bmatrix} x_1(k-1) \\ x_2(k-1) \\ ... \\ x_m(k-1) \end{bmatrix} + B_2 \begin{bmatrix} x_1(k-2) \\ x_2(k-2) \\ ... \\ x_m(k-2) \end{bmatrix} + ... + B_q \begin{bmatrix} x_1(k-q) \\ x_2(k-q) \\ ... \\ x_m(k-q) \end{bmatrix} \quad (5)$$

As it is shown in equation 5, the displacement data corresponding to the endogenous variables at time instant *k* is estimated combining the VARX model and the measured and estimated variables (if it corresponds to the suppressed variable) at previous time instants. This process is carried out repeatedly from *k* equal to *p* until *k* equal to *N+p* in order to estimate *N* samples of the endogenous variables $\hat{y}_1, \hat{y}_2...\hat{y}_n$.

Once the total displacement data have been estimated, a damage indicator (DI) is computed as in equation 6. This damage indicator is obtained as the total mean deviation between the measured endogenous variables and their corresponding estimated values:

$$DI = \frac{\sum_{k=1}^{M}(\sum_{i=1}^{N}|\hat{y}_{ki} - y_{ki}|)}{M*N} \tag{6}$$

*N* represents the amount of estimated samples for each endogenous variable and *M* represents the amount of measured variables. Taking into account the assumptions that we have made during this section, the measured endogenous variables are $(y_2...y_n)$ and their corresponding estimated values are ($\hat{y}_2...\hat{y}_n$), so *M* is equal to *n-1*.

## 3 NUMERICAL RESULTS

A linear and time invariant two-dimensional lattice structure is monitored in this section (see figure 2). The structure consists of stainless steel bars that are connected together by rigid joints and we assume that the forces can only be transmitted along the axial direction of the bars and the load can only be applied at the two ends of each bar. The structural behaviour is described by a lumped parameter model, where we assume that all object are rigid bodies and all interactions between the rigid bodies take place via springs.

As is shown in figure 2, a ten DOFs substructure is selected from the entire structure for its monitoring.

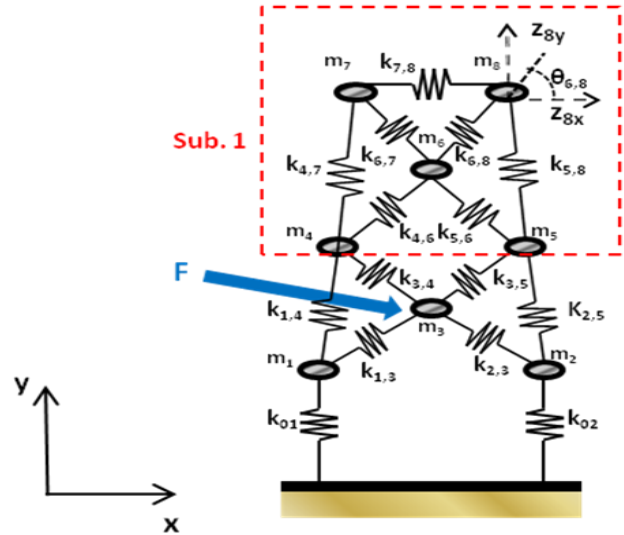

**Figure 2. Isolated substructure in the structural model**

The healthy structure is excited in the third mass (outside the substructure) by a Gaussian white noise and the displacements are measured for each substructure DOF using a data sampling frequency of 1000 Hz during 2 seconds.

By means of the substructuring method explained in [9], the selected substructure is isolated from the general structure. The isolated substructure is represented as a four exogenous and six endogenous variables VARX model. The exogenous variables are the measured absolute displacements in $z_{4x}$, $z_{4y}$, $z_{5x}$ and $z_{5y}$ and the endogenous variables are the measured absolute displacement in $z_{6x}$, $z_{6y}$, $z_{7x}$, $z_{7y}$, $z_{8x}$ and $z_{8y}$. On the other hand, $A_1$ and $A_2$ are 6 x 6 endogenous coefficient matrices and $B_1$ is a 6 x 4 exogenous coefficient matrix. The healthy VARX model is estimated by the Multivariate Least-Square estimator (MLS) [21].

$$\begin{bmatrix} z_{6x}(k) \\ z_{6y}(k) \\ z_{7x}(k) \\ z_{7y}(k) \\ z_{8x}(k) \\ z_{8y}(k) \end{bmatrix} = A_1 \begin{bmatrix} z_{6x}(k-1) \\ z_{6y}(k-1) \\ z_{7x}(k-1) \\ z_{7y}(k-1) \\ z_{8x}(k-1) \\ z_{8y}(k-1) \end{bmatrix} + A_2 \begin{bmatrix} z_{6x}(k-2) \\ z_{6y}(k-2) \\ z_{7x}(k-2) \\ z_{7y}(k-2) \\ z_{8x}(k-2) \\ z_{8y}(k-2) \end{bmatrix} + B_1 \begin{bmatrix} z_{4x}(k-1) \\ z_{4y}(k-1) \\ z_{5x}(k-1) \\ z_{5y}(k-1) \end{bmatrix} \qquad (7)$$

*3.1. Substructure pairwise conditional Granger causality analysis*

The pairwise conditional Granger causality analysis between all measured substructural DOFs (healthy state) is done in this section in order to determine which are the most appropriate ones to be suppressed.

The bidirectional displacement data corresponding to each endogenous node is assigned to an individual data set. The first data set contains displacement data from node 6 ($z_{6x}$, $z_{6y}$), the second data set contains the displacement data from node 7 ($z_{7x}$, $z_{7y}$) and the last data set is formed by displacement data from node 8 ($z_{8x}$, $z_{8y}$).

Once the data sets are defined, an individual pairwise conditional Granger causality analysis is carried out between them as is explained in section 2. As is shown in table 1, six combinations are analysed individually and their respective individual G-causality value ($F_{Y \rightarrow X|Z}$) is calculated.

**Table 1. Substructure's individual pairwise conditional Granger causality analysis**

| X | Y | Z | $F_{Y \rightarrow X|Z}$ |
|---|---|---|---|
| $z_{7x}, z_{7y}$ | $z_{6x}, z_{6y}$ | $z_{8x}, z_{8y}$ | $1.22e^{-2}$ |
| $z_{8x}, z_{8y}$ | $z_{6x}, z_{6y}$ | $z_{7x}, z_{7y}$ | $1.26e^{-2}$ |
| $z_{6x}, z_{6y}$ | $z_{7x}, z_{7y}$ | $z_{8x}, z_{8y}$ | $8.52e^{-3}$ |
| $z_{8x}, z_{8y}$ | $z_{7x}, z_{7y}$ | $z_{6x}, z_{6y}$ | $1.06e^{-2}$ |
| $z_{6x}, z_{6y}$ | $z_{8x}, z_{8y}$ | $z_{7x}, z_{7y}$ | $8.67e^{-3}$ |
| $z_{7x}, z_{7y}$ | $z_{8x}, z_{8y}$ | $z_{6x}, z_{6y}$ | $1.09e^{-2}$ |

The individual G-causality values that are calculated assigning the same data set to *Y* are added together in order to obtain a general causality value ($F_{G\,Y \rightarrow X|Z}$) for each set of displacements. The result is shown in table 2.

**Table 2. Substructure's global pairwise conditional Granger causality analysis**

| Y | $F_{G\,Y \rightarrow X|Z}$ |
|---|---|
| $z_{6x}, z_{6y}$ | $2.480e^{-2}$ |
| $z_{7x}, z_{7y}$ | $1.912e^{-2}$ |
| $z_{8x}, z_{8y}$ | $1.957e^{-2}$ |

The set that contains displacement data of node 6 has the higher global G-causality value, whereas data sets corresponding to nodes 7 and 8 have similar values between them, but significantly lower than the previous case. Therefore, data from node 6 ($z_{6x}$, $z_{6y}$) is the least appropriate to be suppressed.

### 3.2. Substructure damage detection with incomplete observations

Different scenarios (healthy and damaged) are assessed by the developed SHM method. All considered damages are stiffness losses of a specific spring (20%) within the structure. Furthermore, three different damage locations are evaluated. In some of them, the damaged springs are within the substructure ($k_{6,7}$ and $k_{7,8}$) and in the others, they correspond to external springs ($k_{1,3}$).

As in the healthy state, the damaged structures are excited in the third mass (outside the substructure) by a Gaussian white noise. The displacements are measured for all substructural DOFs using a data sampling frequency of 1000 Hz during 2 seconds.

Granger causality analysis has determined that displacement data from nodes 7 or 8, even both, are appropriate to be suppressed. Therefore, the condition of the simulated scenarios is evaluated by the developed damage detection method suppressing firstly data from node 7, secondly data from node 8 and finally displacement data from nodes 7 and 8.

Although a particular case (suppression of data from nodes 7 and 8) is described below, the steps that must be followed are equals in the other two cases. These steps are summarised in figure 1.

The displacement data of the endogenous variables ($\hat{z}_{6x}, \hat{z}_{6y}, \hat{z}_{7x}, \hat{z}_{7y}, \hat{z}_{8x}, \hat{z}_{8y}$) at time instant $k$ is estimated combining the healthy VARX model ($A_1, A_2, B_1$) and the measured ($z_{4x}, z_{4y}, z_{5x}, z_{5y}, z_{6x}, z_{6y}$) and estimated ($\hat{z}_{7x}, \hat{z}_{7y}, \hat{z}_{8x}, \hat{z}_{8y}$) variables at previous time instants. This process is carried out repeatedly from $k$ equal to $2$ until $2002$ in order to estimate $2000$ samples of the endogenous variables.

$$\begin{bmatrix} \hat{z}_{6x}(k) \\ \hat{z}_{6y}(k) \\ \hat{z}_{7x}(k) \\ \hat{z}_{7y}(k) \\ \hat{z}_{8x}(k) \\ \hat{z}_{8y}(k) \end{bmatrix} = -A_1 \begin{bmatrix} z_{6x}(k-1) \\ z_{6y}(k-1) \\ \hat{z}_{7x}(k-1) \\ \hat{z}_{7y}(k-1) \\ \hat{z}_{8x}(k-1) \\ \hat{z}_{8y}(k-1) \end{bmatrix} - A_2 \begin{bmatrix} z_{6x}(k-2) \\ z_{6y}(k-2) \\ \hat{z}_{7x}(k-2) \\ \hat{z}_{7y}(k-2) \\ \hat{z}_{8x}(k-2) \\ \hat{z}_{8y}(k-2) \end{bmatrix} + B_1 \begin{bmatrix} z_{4x}(k-1) \\ z_{4y}(k-1) \\ z_{5x}(k-1) \\ z_{5y}(k-1) \end{bmatrix} \qquad (8)$$

Once the total displacement data ($\hat{z}_{6x}, \hat{z}_{6y}, \hat{z}_{7x}, \hat{z}_{7y}, \hat{z}_{8x}, \hat{z}_{8y}$) have been estimated, a damage indicator (DI) is computed as in equation 6. This damage indicator is obtained as the total mean deviation between the measured endogenous variables ($z_{6x}, z_{6y}$) and their corresponding estimated values ($\hat{z}_{6x}, \hat{z}_{6y}$).

Figure 3 shows the obtained results for three kinds of scenarios; healthy (a), damage out of the substructure (b) and damage within substructure (c). Blue and green bars represent the DI values when data from node 7 or node 8 is suppressed respectively, whereas the results obtained suppressing the data from both nodes is depicted by brown bars.

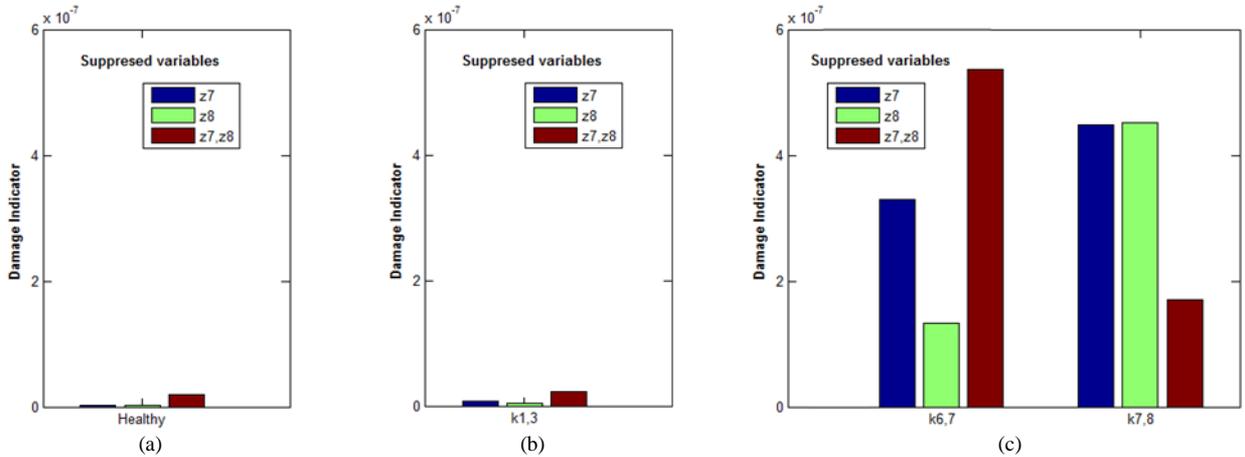

**Figure 3. Calculated DI values for the simulated scenarios**

Regarding to the results, when the whole structure is healthy or when the damage is introduced out of the substructure (reducing $k_{1,3}$), the calculated DI values are always almost zero.

On the other hand, if the damages are introduced within the substructure ($k_{6,7}$ and $k_{7,8}$) the DI values are much higher than in the previous cases. Besides, the mentioned conclusions are valid even thought data from the two nodes is suppressed.

Figure 3 shows that the DI values are practically the same if $k_{7,8}$ is reduced and $z_7$ and $z_8$ are suppressed individually. However, if $k_{6,7}$ is modified, the corresponding DI values are different. This fact is related to the position of the damaged element, because $k_{7,8}$ correspond to a spring situated between node 7 and node 8, whereas $k_{6,7}$ has more influence in node 7 than in node 8.

## 4 Conclusions

This paper proposes a SHM method to detect damages within substructures using incomplete measurements. In the characterization step, which corresponds to the healthy state, the displacements from all substructural nodes are measured and the corresponding VARX model is estimated. Furthermore, pairwise conditional Granger causality analysis is carried out with the measured data in order to select which displacement data is the most appropriate to be suppressed. In the evaluation phase, the healthy VARX model and the displacement data measured from the selected nodes are used to determine the condition of the substructure.

A linear and time invariant model of a two-dimensional lattice structure is simulated to validate the proposed method. The results show that the method allows detecting damages within substructures using a unique accelerometer.

The proposed method is also suited for three dimensional lattice structures, where the number of element's connections increases. Our research group is already applying this method in a real structure and the results will be published soon.

## 5 Acknowledgement

This study is partially funded by the Ministry of Economy and Competitiveness of the Spanish Government (WINTURCON project, DPI2014-58427-C2-2-R). Any opinions, findings and conclusions expressed in this article are those of the authors and do not necessarily reflect the views of the funding agency.